\pdfoutput=1

\documentclass[final,5p,times,twocolumn]{elsarticle}

\usepackage{graphicx}

\usepackage{amssymb}

\usepackage{lineno}

\biboptions{comma,square}

\journal{Nuclear Intruments and Methods in Physics Research A}

\DeclareMathSymbol{,}{\mathpunct}{letters}{"3B}
\DeclareMathSymbol{.}{\mathord}{letters}{"3B}
\DeclareMathSymbol{\decimal}{\mathord}{letters}{"3A}

\begin{document}

\begin{frontmatter}



\title{IceCube: physics, status, and future }


\author{Klas Hultqvist, for the IceCube collaboration}

\address{Oskar Klein Centre, Dept. of Physics, Stockholm University, 106 91 Stockholm, Sweden}

\begin{abstract}
The IceCube observatory is the first cubic kilometre scale instrument in
the field of high-energy neutrino astronomy and cosmic rays.
In 2009, following five successful deployment seasons, IceCube
consisted of 59 strings of optical modules in the
South Pole ice, together with 118 air shower detectors in the IceTop
surface array. The range of physics topics includes 
neutrino signals from astrophysical sources, dark matter, exotic
particle physics, cosmic rays, and atmospheric neutrinos. 
The current IceCube status and selected results are described. 
Anticipated future developments 
are also discussed, in particular the Deep Core low energy 
subarray which was recently deployed.
\end{abstract}


\end{frontmatter}


\section{Introduction}
\label{sec:intro}
At the IceCube observatory, the clear South Pole glacial ice is exploited 
to detect Cherenkov light emitted by charged particles
created in neutrino interactions. An array of digital optical modules
(DOMs) equipped with photomultipliers extends throughout a volume of 
approximately $1\,{\rm km}^3$, with the
central region more densely intrumented. Scheduled for completion in
2011, IceCube allows the detection of neutrino interactions above
a few tens of GeV, with a maximum sensitivity in the TeV-PeV range.

Extraterrestrial neutrinos are expected from a variety of sources. 
Generically, high energy cosmic ray particles will give rise to 
neutrinos in their interactions with matter or radiation, either in the 
immediate vicinity of the acceleration sites or while traveling through 
space. It is thus possible to estimate the neutrino flux that corresponds 
to the observed flux of cosmic rays~\cite{WBbound}. Such considerations have 
dictated sensitivity requirements which translate into a cubic kilometre 
detector. 

Owing to their low interaction cross-section and lack of 
electric charge, neutrinos propagate over vast distances, quite undisturbed 
by matter, radiation, and magnetic fields. This is in contrast to photons,
which can be absorbed by matter, and which cannot travel cosmological
distances if their energy is above the TeV range, because of absorption by
background radiation. Similarly, protons of the highest energies can be
absorbed through the GZK mechanism~\cite{GZK}, while at lower energies 
they are deflected by magnetic fields and do not point back to the source.
It is therefore possible that there are ``hidden'' sources which can 
only be detected
via their neutrino emission. In cases where emission of gamma rays
or high energy cosmic rays is seen, a detection of neutrinos could
elucidate the nature and location of cosmic ray accelerators. (Gamma
rays from $\pi^0$ decay would be accompanied by neutrinos from $\pi^+/\pi^-$
decay.)
Promising candidate acceleration sites are relativistic ejecta from 
active galactic nuclei (AGNs) or gamma-ray bursts (GRBs). There are 
also objects within our galaxy, such as supernova remnants, pulsar winds, 
or microquasars, in which acceleration processes could yield neutrino emission.

Even if individual neutrino sources are not seen, their combined 
{\em diffuse} flux may stand out above the atmospheric 
neutrino flux. The study of the atmospheric neutrino flux is therefore an 
important topic. It provides a useful calibration point and may in addition 
yield results relevant to the intrinsic properties of neutrinos.

There are abundant indications of dark
matter in the form of non-relativistic weakly 
interacting massive particles (WIMPs)~\cite{DMrev}. These could
constitute a class of ``hidden'' sources, detectable via neutrinos, 
as they annihilate in the centre of the Sun or other dense regions
~\cite{WIMPcapture,WIMPsusy}.
Promising candidate WIMPs are the neutralinos of supersymmetric 
models~\cite{WIMPsusy} or Kaluza-Klein excitations in models of 
extra dimensions~\cite{HooperProfumoUED}.

Section \ref{sec:detstat} describes the IceCube detector and its
performance, and selected analysis results are discussed in
section \ref{sec:results}. 
Future enhancements are discussed in section \ref{sec:future}.

\section{The IceCube detector}
\label{sec:detstat}

\begin{figure}
\begin{center}
\includegraphics[width=7cm]{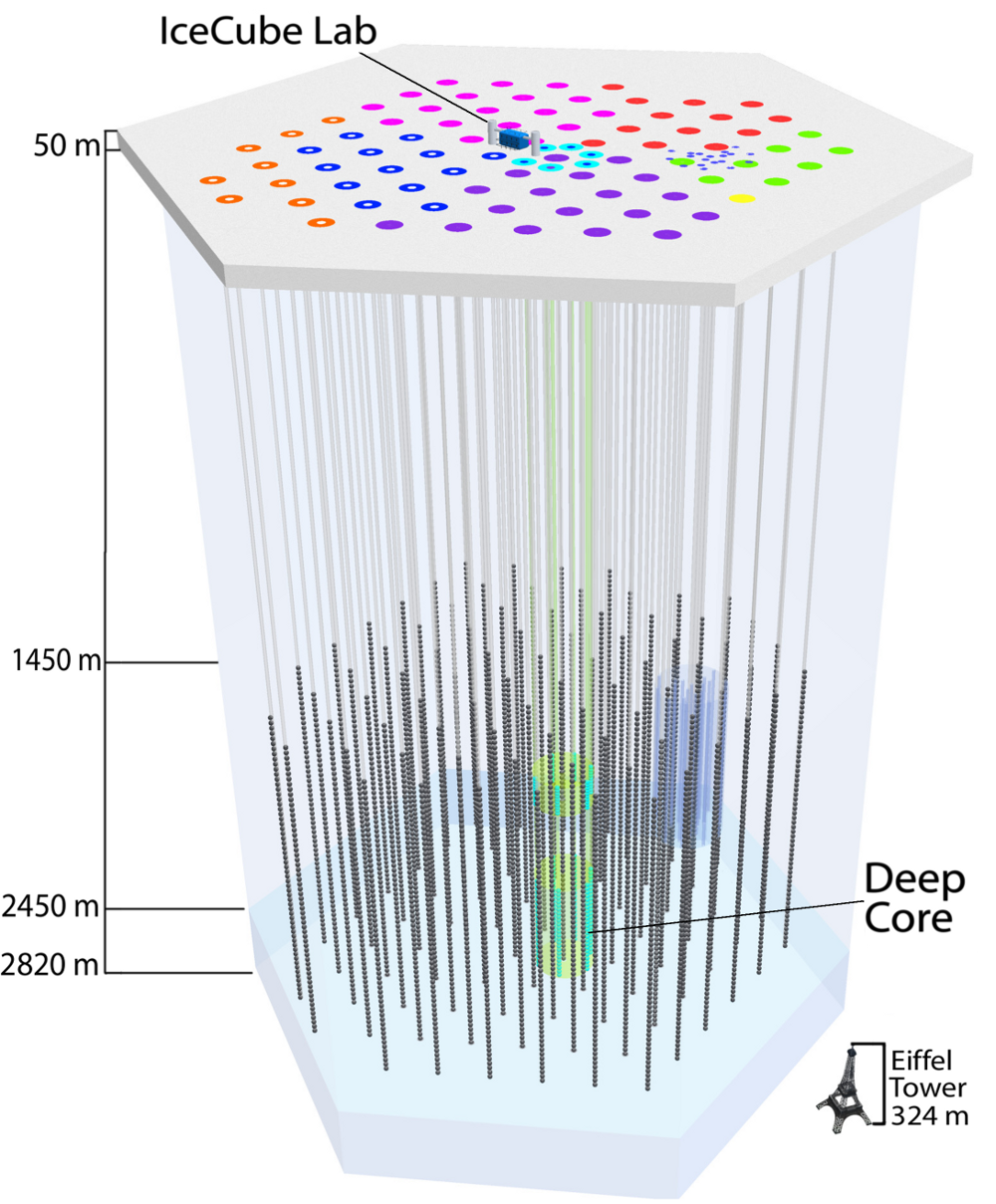}
\end{center}
\caption{The IceCube detector. Filled circles indicate the 59
strings completed for the 2009 data taking season. The shaded region near
the centre is the Deep Core subarray, and the one to the right shows
the position of AMANDA.}
\label{fig:detview}
\end{figure}

Figure \ref{fig:detview} illustrates the layout of the IceCube observatory
in the polar icesheet. Also shown is the AMANDA
array~\cite{AMANDA}, which continued to
operate as a part of IceCube until it was finally closed 
down in May 2009.
In addition to the deep ice detector, there is a surface array, IceTop, 
for detecting cosmic ray air showers. Each IceTop station is located 
above a string, and consists of two tanks. 
Each tank is filled with clear ice and monitored by two DOMs running
at different gains to achieve the required 
dynamic range.

IceCube construction takes place during the austral summer. A 5 MW hot-water 
drilling system is used to melt 
holes in the ice down to a depth of $2.45\; {\rm km}$. In each hole
a string consisting of a cable carrying 60 optical modules 
is deployed, and the water in the hole then re-freezes. During the recent
deployment season, the time between completed deployments was about two days. 
When completed in 2011, IceCube will consist of at least 80 strings and
80 IceTop stations. Fifty-nine strings were in operation during 2009, and 
an additional 20 were deployed in the summer season of 2009/2010.

\begin{figure}
\includegraphics[width=7cm]{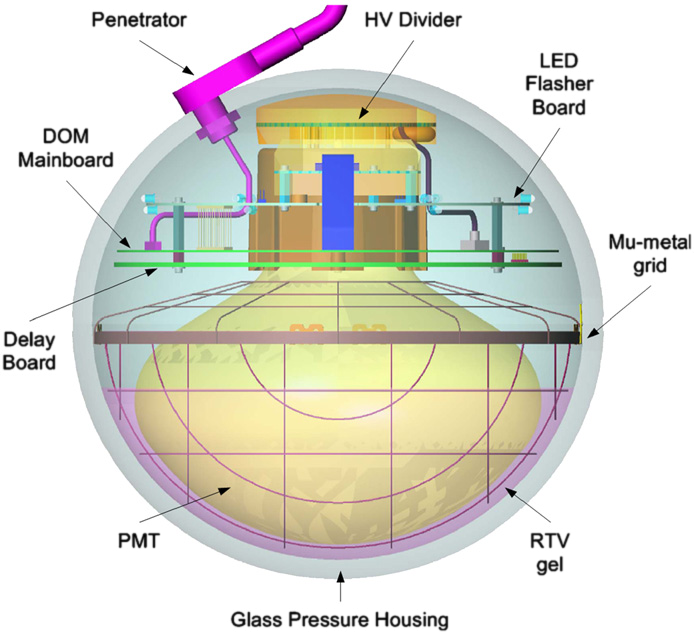}
\caption{A schematic drawing of an IceCube optical module (DOM).
Figure from reference\citealt{DOMpaper}.}
\label{fig:DOM}
\end{figure}

Figure \ref{fig:DOM} shows an IceCube optical module~\cite{DOMpaper}. 
The light-sensitive element is a $25\, {\rm cm}$ photomultiplier 
tube (PMT), looking down. This is optically
coupled to a surrounding pressure sphere via transparent gel. The
sphere also contains a high-voltage generator, 12 LED flashers for
calibration purposes, communication and time-calibration electronics,
and two different types of digitization hardware. Three Analogue
Transient Waveform Digitizer (ATWD) channels with different gains are used
to digitize the PMT output with nanosecond precision during $400\, {\rm ns}$. 
There is also a fast ADC running at $40\, {\rm MHz}$ 
during $6.4\, \mu{\rm s}$. The DOM is triggered by a discriminator coupled
directly to the PMT output, and the input to the digitizers is delayed 
by $75\, {\rm ns}$ to allow the capture of the complete waveform.

An important component in the IceCube detector is the ice itself.
Its optical properties have been measured in situ using light 
emitting devices~\cite{IceMeasmt}, and the ice model and its 
implementation 
in light propagation and reconstruction software are being
gradually refined. Scattering dominates over absorption, with 
an overall attenuation length in excess of 20 m in the upper
regions of the detector, and about twice that value below
a depth of 2100 m. In between these two regions there is 
a dusty layer with strong scattering and absorption. 

IceCube is designed to detect neutrinos of all flavours and distinguish
extraterrestrial neutrino signals from the background. The remainder of
this section contains a discussion of the principles involved, as well 
as some examples of detector performance. In all cases, detection relies 
on the emission of Cherenkov light, at an angle of 
$41^\circ$ from the direction of propagation of charged relativistic 
particles. 

\begin{figure}[!ht]
\includegraphics[width=10cm]{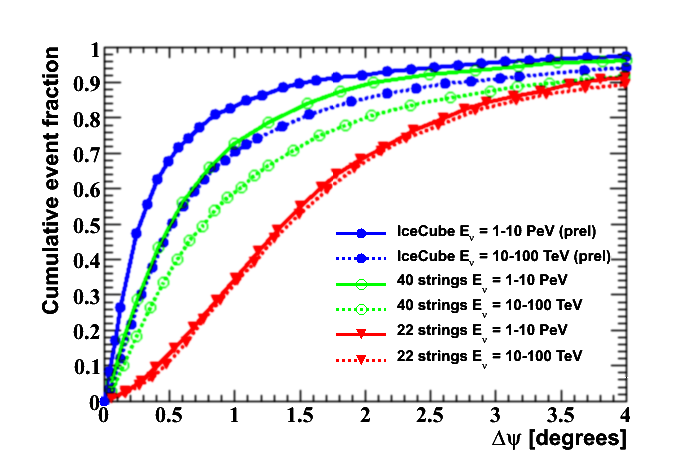}
\caption{The cumulative point spread function for well reconstructed
events, including the angle between the incoming
neutrino and the muon. The improvement between the 22- and 40 string
case is not only due to the larger detector, but also to an improved
reconstruction algorithm. Figure from reference\citealt{IC40ptsrc}.}
\label{fig:ptspread}
\end{figure}

When a muon from
a charged current (CC) $\nu_\mu$ interaction passes through the instrumented
volume it gives rise to a characteristic ``track'' consisting of large 
pulses in DOMs close to the muon trajectory, separated in time by 
the muon flight time. Its direction can therefore be well determined.
The point spread function for well reconstructed muon neutrino events is
shown in Figure \ref{fig:ptspread}. Sub-degree resolution has been 
achieved already with the 40-string configuration.
Muons which pass close to the edge of the instrumented volume, or
outside, can also be detected, especially at high energies. In this case,
however, the event reconstruction and background rejection is more
difficult, and relies more heavily on the description of the
emission of Cherenkov light and its propagation through the ice.

\begin{figure}[!ht]
\includegraphics[width=9cm]{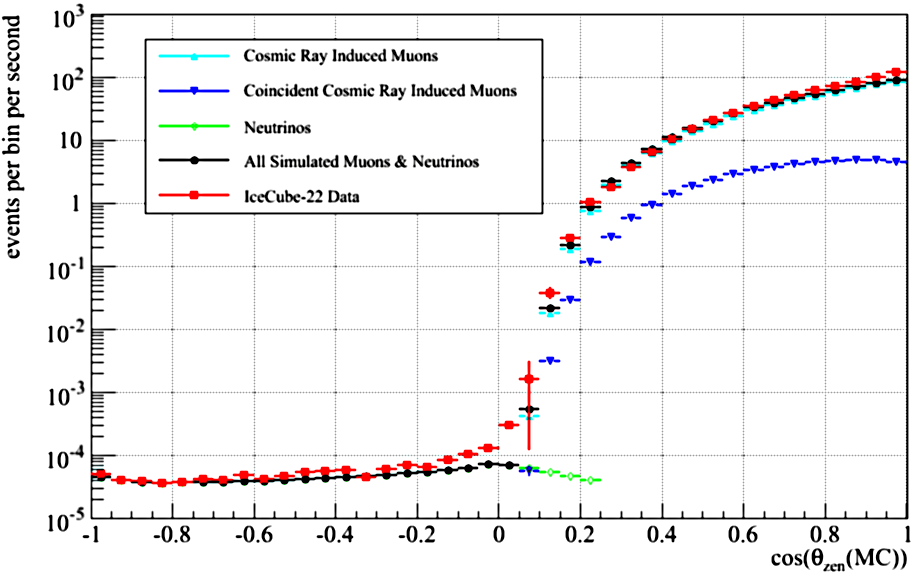}
\caption{
Muon trigger rates in the 22-string IceCube configuration,
as a function of the cosine of the zenith angle. The rate
in data (red squares) is compared to a simulation including both
atmospheric muons and muons from interacting atmospheric
neutrinos.
Requirements to select
well reconstructed events were applied, and the rates are
corrected for the efficiency of these.
Figure from reference\citealt{BerghausISVHECRI}.
\label{fig:musandnus}
}
\end{figure}

To achieve sensitivity
to extraterrestrial neutrino sources, one must also
eliminate the overwhelming background of
downgoing muons from cosmic ray air showers.
The traditional approach is to use the Earth as a filter
and select up-going events. There
will still be a large background from downgoing muons which
are mis-reconstructed as upgoing. By requiring well reconstructed
events, however, one can define a clean sample of 
up-going neutrinos, as illustrated in Figure \ref{fig:musandnus}.

\begin{figure}
\includegraphics[width=9cm]{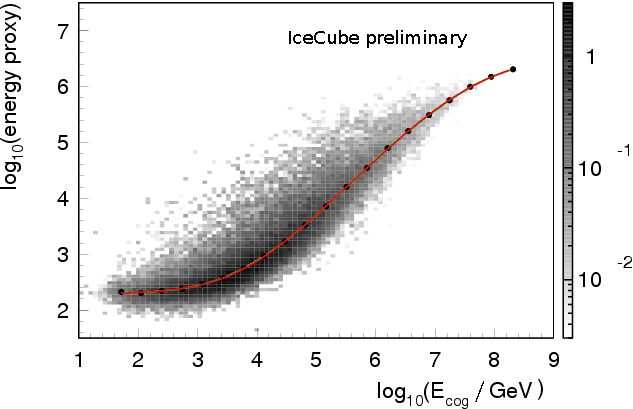}
\caption{
Reconstructed vs true logarithm of the muon energy, at the point 
of closest approach to the centre-of-gravity of hit modules. Note 
the logarithmic intensity scale.
\label{fig:mue}
}
\end{figure}

The energy spectrum of atmospheric neutrinos falls rapidly, as
$\sim E^{-3.7}$, whereas neutrinos from atrophysical sources
are expected to have a much harder spectrum ($\sim E^{-2}$). 
An energy estimator
will therefore provide additional discrimination in the search for
extraterrestrial point sources. Furthermore, it is essential in the 
search for a diffuse signal of extraterrestrial muon neutrinos.
For muons above a few TeV, 
the amount of Cherenkov light scales roughly linearly with energy, and
the total number of triggered DOMs has been used as an 
energy proxy. Recently, a more direct method for measuring
the energy loss along the muon trajectory has been 
developed~\cite{ChirkinEmuICRC09,KarleICRC09}, 
with a resolution in
$\log_{10} (E_\mu / {\rm GeV})$ of around 30\% in the range between
1 TeV and 100 PeV, see Figure \ref{fig:mue}.

\begin{figure}
\includegraphics[width=9cm]{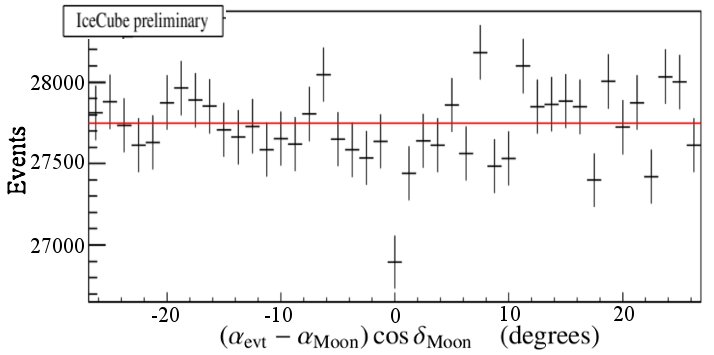}
\caption{
Angular distance from the Moon for events in a $1.25^\circ$ declination 
band centered on the Moon.
A $5.2 \sigma$ deficit is seen in the central bin. The figure is
based on eight months of data taken with the 40-string 
configuration in 2008.
\label{fig:moonshadow}
}
\end{figure}

Two recent observations of downgoing atmospheric muons illustrate that 
the reconstruction methods work. One is the detection of the 
reduction of the cosmic ray flux caused by 
absorption in the Moon~\cite{moonshadow}, see Figure \ref{fig:moonshadow}.
Systematic biases in the reconstruction could affect the position of the
minimum. No such effect is seen in the analysed data. Increased statistics 
will render the test more stringent.
\begin{figure}
\includegraphics[width=9.5cm]{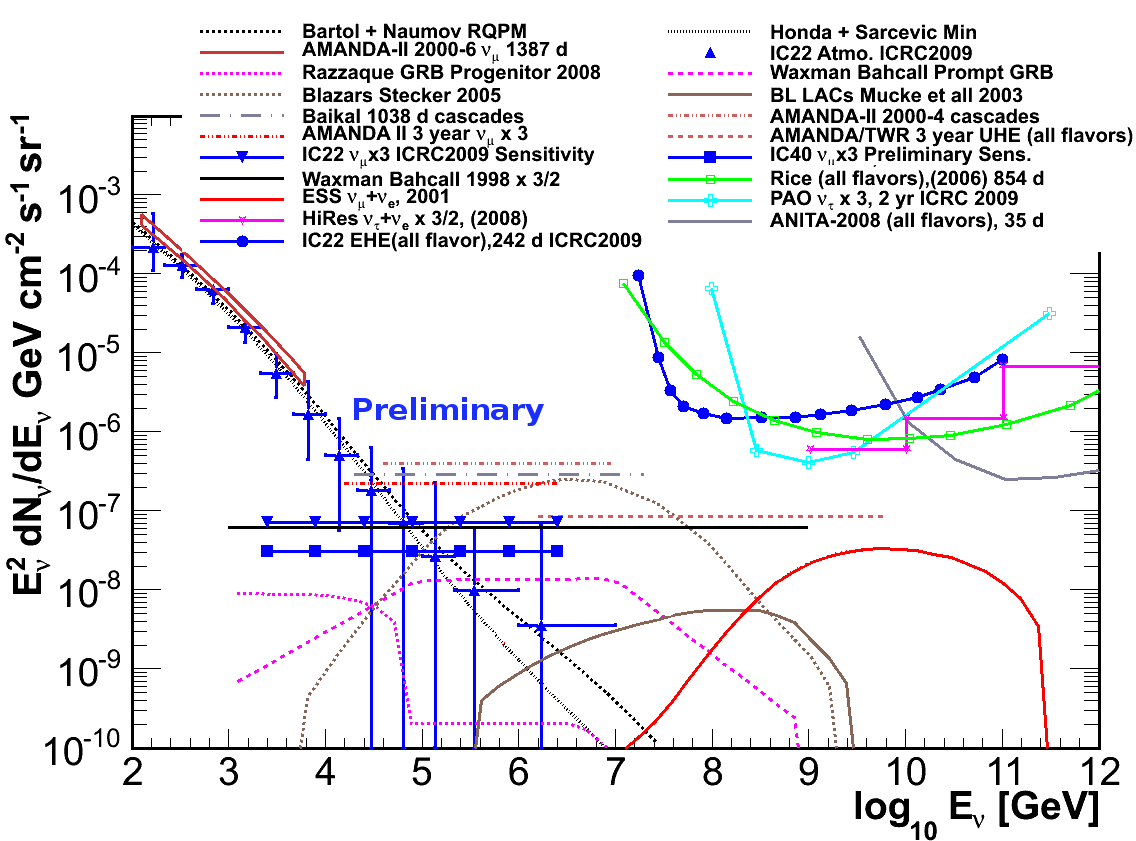}
\caption{
Diffuse neutrino flux predictions, measurements, and limits. All results
are scaled to apply to the total neutrino flux, assuming that the
fluxes are equal for all three flavours. See text for details.
\label{fig:diffuseflux}
}
\end{figure}
The other result is the observation of a global anisotropy in the
cosmic ray arrival directions~\cite{CRanis}. This is the first
such measurement in the Southern Hemisphere, and it agrees well with
measurements done in the North~\cite{TIBETanis,MILAGROanis}.

Neutral current interactions give cascade events in which no
particle travels a distance comparable to the spacing between 
detector modules. This is also true for $\nu_{\rm e}$ CC events below the
EeV range, and for $\nu_\tau$ CC events below a few PeV. 
For such cascades, the light is emitted from a small region, but 
preferentially at the Cherenkov angle of $41^\circ$ with respect to 
the shower axis.
Making use of this information, and detailed knowledge about light
propagation in the ice, an angular resolution of 30 to 35 degrees
is expected for cascades in the energy range 
between 10 TeV and 10 PeV~\cite{ImpCascRec}. The resolution
in $\log_{10} (E/{\rm GeV})$ is $\sim 13\%$ for cascades contained in the
instrumented volume~\cite{ImpCascRec}. As the array grows, the 
volume where a cascade can be identified increases faster 
than the detector volume, and much faster than 
the area exposed to a throughgoing muon. Therefore the cascade 
channel, with its better energy resolution, is becoming more 
important as the detector grows.
 
At higher energies electromagnetic showers initiated by electrons from
$\nu_{\rm e}$ interactions become elongated owing to the
LPM effect~\cite{LPM,KleinLPM}, which makes a better angular reconstruction 
possible~\cite{BolmontLPM}. For $\nu_\tau$ energies in the PeV range and
above, the
distance travelled by the $\tau$ from a CC interaction leads to several
signatures~\cite{CowenTEVPA2}. These are of special interest 
as there is no terrestrial $\nu_\tau$ background flux, whereas 
astrophysical neutrino fluxes are expected to contain a mix of all 
flavours.

\section{Selected results \label{sec:results}} 

Figure \ref{fig:diffuseflux} shows two preliminary results on the diffuse 
neutrino flux, obtained from the IceCube data taken in 2007 with 22 
strings (IC22). 
The triangles with error bars show a preliminary energy spectrum, obtained
using an unfolding method~\cite{ChirkinEmuICRC09}. Agreement is found with
previous AMANDA unfolding results, and with the calculated atmospheric
neutrino flux.
The connected filled circles show a quasi-differential 
limit\footnote{This is the limit on a spectrum which falls as
$E^{-2}$ over one decade in energy, and is zero outside this range.}
at 90\% confidence level, obtained in a preliminary low background 
search for Extremely High Energy (EHE) events~\cite{EHEsearch}. 
No candidates were found.

Also shown are projected sensitivities of the searches with IC22 and 
IC40 (the 2008 40-string configuration) for extraterrestrial
diffuse fluxes falling as $E^{-2}$ \cite{KarleICRC09}. 
They are drawn as constant lines 
with a length corresponding to a region which would contain 90\% of the 
selected signal events. 
The figure also shows results obtained with AMANDA. These are an unfolded
energy spectrum for atmospheric neutrinos, based on seven years of
muon data~\cite{AMANDAespec}, limits on a diffuse $E^{-2}$ flux obtained
using muons~\cite{AMANDAmulim} and cascades~\cite{AMANDAcasclim}, and
a limit on the flux of Ultra High Energy (UHE) neutrinos based on three
years of AMANDA data~\cite{AMANDAuhelim}.

Figure \ref{fig:diffuseflux} also includes some results from
other detectors.
They are, in order of increasing energy,
the Lake Baikal neutrino observatory,
the RICE South Pole radio Cherenkov array, the Pierre Auger Observatory 
(PAO),
the HiRes air fluorescence experiment, and the ANITA radio Cherenkov
detection antenna.
Also shown are 
the Waxman-Bahcall flux derived from the observed cosmic ray 
spectrum~\cite{WBbound} and calculated contributions
to the diffuse neutrino fluxes from atmospheric neutrinos, 
gamma-ray bursts~\cite{RazzaqueWB}, 
active galactic nuclei~\cite{SteckerMucke}, and the GZK process~\cite{ESS}.

Some of the neutrino 
sources contributing to a diffuse flux may be 
individually seen as point sources. Since the energy spectrum of the
signal is expected to be harder than that of the atmospheric neutrino
background, a likelihood ratio hypothesis test including 
energy and direction is the natural approach~\cite{BraunMethod}. 
Repeating this for a very fine grid of source directions gives a map of 
the $p$-value (the a priori probability, in the absence of a source,
to obtain a result which is at least as signal-like as 
the one actually obtained). 

\begin{figure}
\includegraphics[width=9cm]{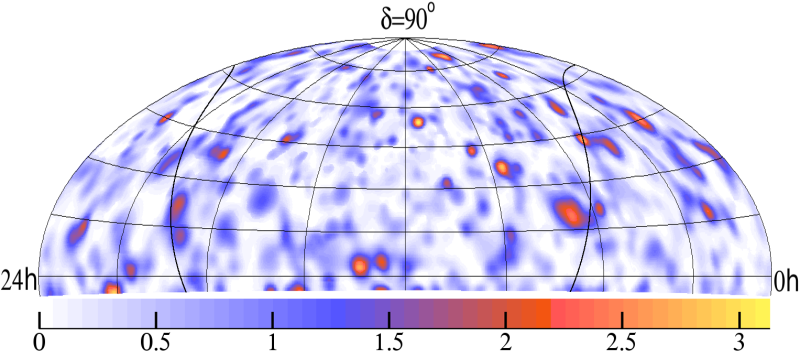}
\caption{
Significance map showing $-\log_{10} p$ for points in the northern sky, 
based on a search for point sources in seven years of AMANDA data.
\label{fig:AMANDA7yrMap}
}
\end{figure}

Figure \ref{fig:AMANDA7yrMap} shows a sky map obtained from seven years of
AMANDA observations~\cite{AMANDA7yr}. The maximum significance is at
$\delta = 54^\circ$, $\alpha = 171^\circ$, where the $p$-value is
$p=7.4 \cdot 10^{-4}$.
The probability of obtaining such a $p$-value or lower somewhere in the sky,
determined by randomising the right ascension, is $p=95\%$. Hence there is
no indication of a point source signal.

The point source search in the data taken with the IceCube 22-string 
configuration~\cite{IC22ptsrc} 
was about twice as sensitive as the AMANDA one, and showed a 
conspicuous deviation from background with 
$p=7 \cdot 10^{-7}$ at $\delta = 11.4^\circ,\,\alpha = 153.4^\circ$. 
However, the final $p$-value, accounting for the multiple trials in 
scanning the sky, is $p=1.3 \%$, which is a very marginal significance.

\begin{figure}[!ht]
\begin{center}
\includegraphics[width=9.3cm]{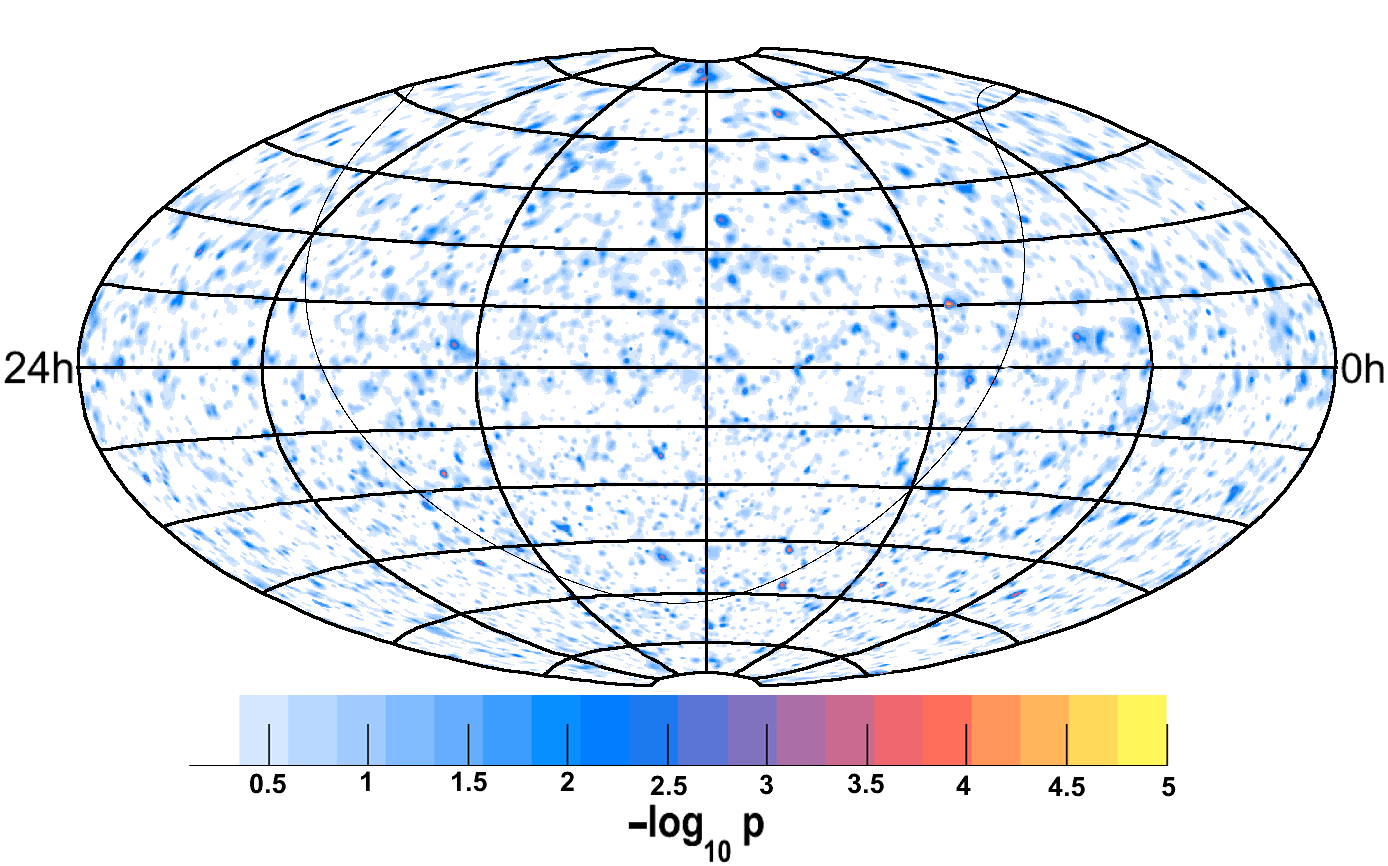}
\end{center}
\caption{
Significance map for a point source search in six months of data taken
with the IceCube 40-string configuration.
There are 6796 events from the northern celestial hemisphere, and 10981 from
the southern one, where the atmospheric muon background dominates.
\label{fig:IC40ptsrc}
}
\end{figure}
A point source analysis of six months of data taken with 40 
strings (out of 13 months in total)~\cite{IC40ptsrc} yielded the sky map
shown in Figure \ref{fig:IC40ptsrc}. 
Here the most significant point
is at $\delta = 15.4^\circ,\,\alpha = 115.0^\circ$ and the 
$p$-value, accounting for multiple trials,
is $61 \%$. 
The map includes the southern sky, 
where the background of downgoing muons completely dominates 
the atmospheric neutrinos. 
A zenith dependent cut on the muon energy parameter
was imposed to ensure a nearly uniform zenith distribution of the background.
As a result, the search in the southern hemisphere is sensitive primarily to
neutrino energies in the PeV range, as compared to TeV energies for the 
northern hemisphere. 
The better angular resolution for higher energy can be 
clearly seen in the figure. 
A search above the horizon has also
been performed using the 22-string data~\cite{IC22horizon}.

In each of the point source analyses mentioned above, source
candidates on a pre-defined list have been tested, in addition
to the very fine grid search. The most significant result is for 
the Geminga pulsar in the seven-year AMANDA data ($p = 0.9\%$). For the
IceCube searches, however, Geminga gives downward fluctuations ($p > 0.5$). The
point showing the excess in the 22-string data was included as a 
candidate in the 40-string search, where it showed a downward
fluctuation. 

A search for time clustering of events from candidate sources has
also been carried out using the 22-string data~\cite{IC22flares}, and
there have been efforts to correlate IceCube observations in time with flares
observed in the electromagnetic spectrum~\cite{IC22PtSrcTimes}.
There is also a programme to provide online triggers for air Cherenkov 
telescopes~\cite{ICMAGIC}, and for fast optical follow-up
observations of transient neutrino emission, as expected from e.g. supernovae or 
gamma-ray bursts~\cite{ICROTSE}.

\begin{figure}
\includegraphics[width=9cm]{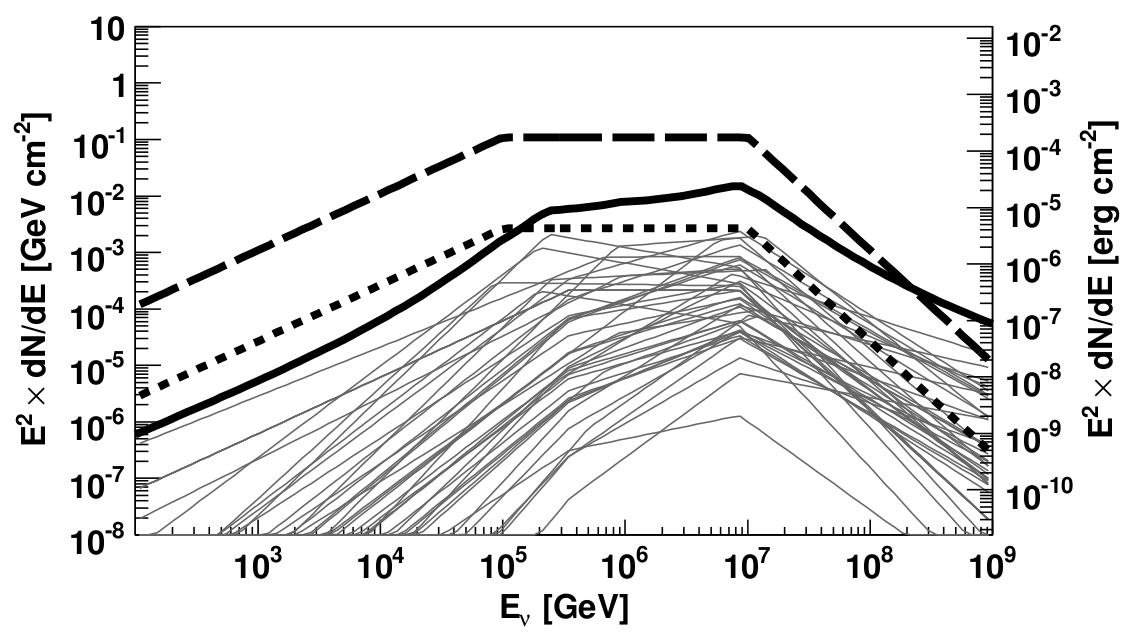}
\caption{
Calculated prompt neutrino fluences for 41 GRBs searched for neutrino 
emission in the 22-string data. Also shown is the sum over all 
bursts (thick solid line), 
the Waxman-Bahcall spectrum~\cite{WBGRBs} for a single burst (thick 
dotted) and for 41 bursts (thick dashed).
\label{fig:41bursts}
}
\end{figure}

In addition, the recorded data are searched 
for neutrino events coincident in time and direction with GRBs detected by
satellites~\cite{GRBmuons,GRBcasc,GRBAMANDA}. The approach has recently 
been refined to model the emission of individual bursts based on the observed 
parameters, as shown in Figure \ref{fig:41bursts}.
Limits obtained~\cite{GRBmuons} are two orders of
magnitude above the fluence sum in Figure \ref{fig:41bursts}, and about a
factor five above a predicted lower energy precursor fluence\cite{RazzaqueWB}. 
A study was also 
made~\cite{GRBmuons} of
the expected sensitivity of the full IceCube detector, with future GRB 
samples from the Fermi and Swift satellites. 
It indicates that fluences, predicted from the assumption that
GRBs are the dominant sources of ultra-high energy cosmic rays,
will be detected or ruled out after some years of running 
IceCube~\cite{GRBmuons}.

In conclusion, no evidence for steady or transient neutrino point sources 
has been found so far.

The indirect detection of dark matter via neutrinos from WIMP annihilations is
another important aim of IceCube. The nearest candidate sites for
gravitational trapping are the centres of the Earth and of the Sun. Since the
light elements dominate the Sun, the spin-dependent scattering cross-section
is more relevant for WIMP capture in the Sun than in the Earth. For the same 
reason, searches for an indirect signal from capture in the Sun complement
those in direct detection experiments, which are primarily sensitive to the
spin-independent cross-section. 

\begin{figure}
\mbox{$\phantom{l}$} 
\includegraphics[width=9cm]{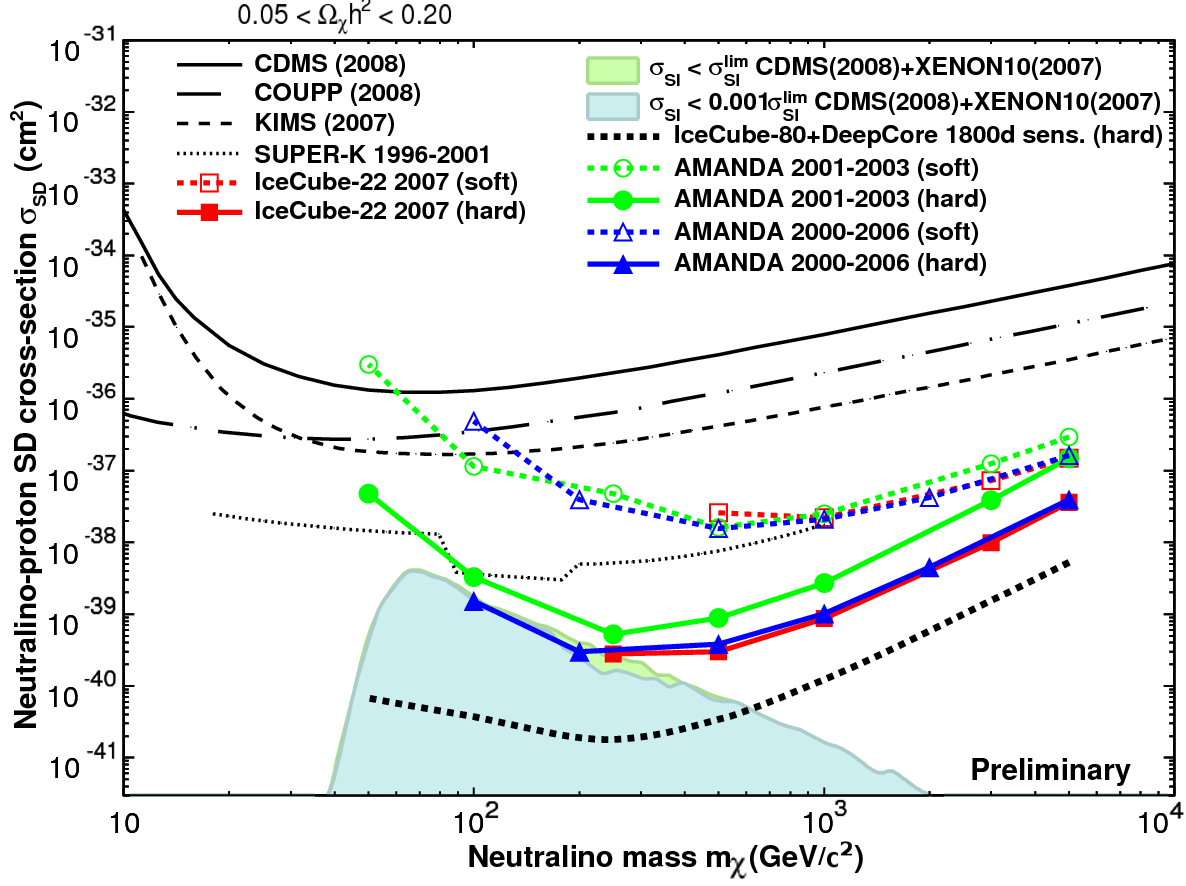}
\caption{
Limits on the spin-dependent neutralino cross-section obtained with IceCube
and AMANDA compared to results from Super-Kamiokande and direct 
detection experiments (cited in reference\citealt{IC22neutralinos}). 
Also shown is the region corresponding to allowed supersymmetric
models found using DarkSusy~\cite{DarkSusy}
for current limits on the spin-independent cross-section, and
the region which would still be allowed if these limits improved 
by a factor 1000.
\label{fig:NeutralinoSigma}
}
\end{figure}

The 22-string IceCube data taken with the Sun below the horizon have been 
analysed, using a likelihood ratio method involving the angle relative 
to the Sun, to test hypotheses of neutralino 
annihilations~\cite{IC22neutralinos}. Neutralino masses in the
range from $250\, {\rm GeV}/c^2$ to $5\, {\rm TeV}/c^2$ were selected, and 
two extreme annihilation channels were investigated, 
the ``hard'' (${\rm W}^+{\rm W}^-$) and 
the ``soft'' (${\rm b}\overline{\rm b}$). The data show a slight
deficit in the direction of the Sun~\cite{IC22neutralinos}.
Following the procedure of reference\citealt{SigmaCalc}, 
the resulting limits on the muon flux in the ice were 
converted into the limits on the spin-dependent cross-section shown in 
Figure \ref{fig:NeutralinoSigma}.
The figure also shows limits obtained using AMANDA data~\cite{AmandaICRCwimps}.
One set was obtained using the data sample and
techniques from the seven-year point source search~\cite{AMANDA7yr}, and 
another from a dedicated analysis of three years of data~\cite{HubertThesis}.
It can also be seen from Figure \ref{fig:NeutralinoSigma} that the 
limits on the spin-independent 
cross-section from direct detection searches do not appreciably constrain 
the supersymmetric model space in this projection.

\begin{figure}
\includegraphics[width=8.5cm]{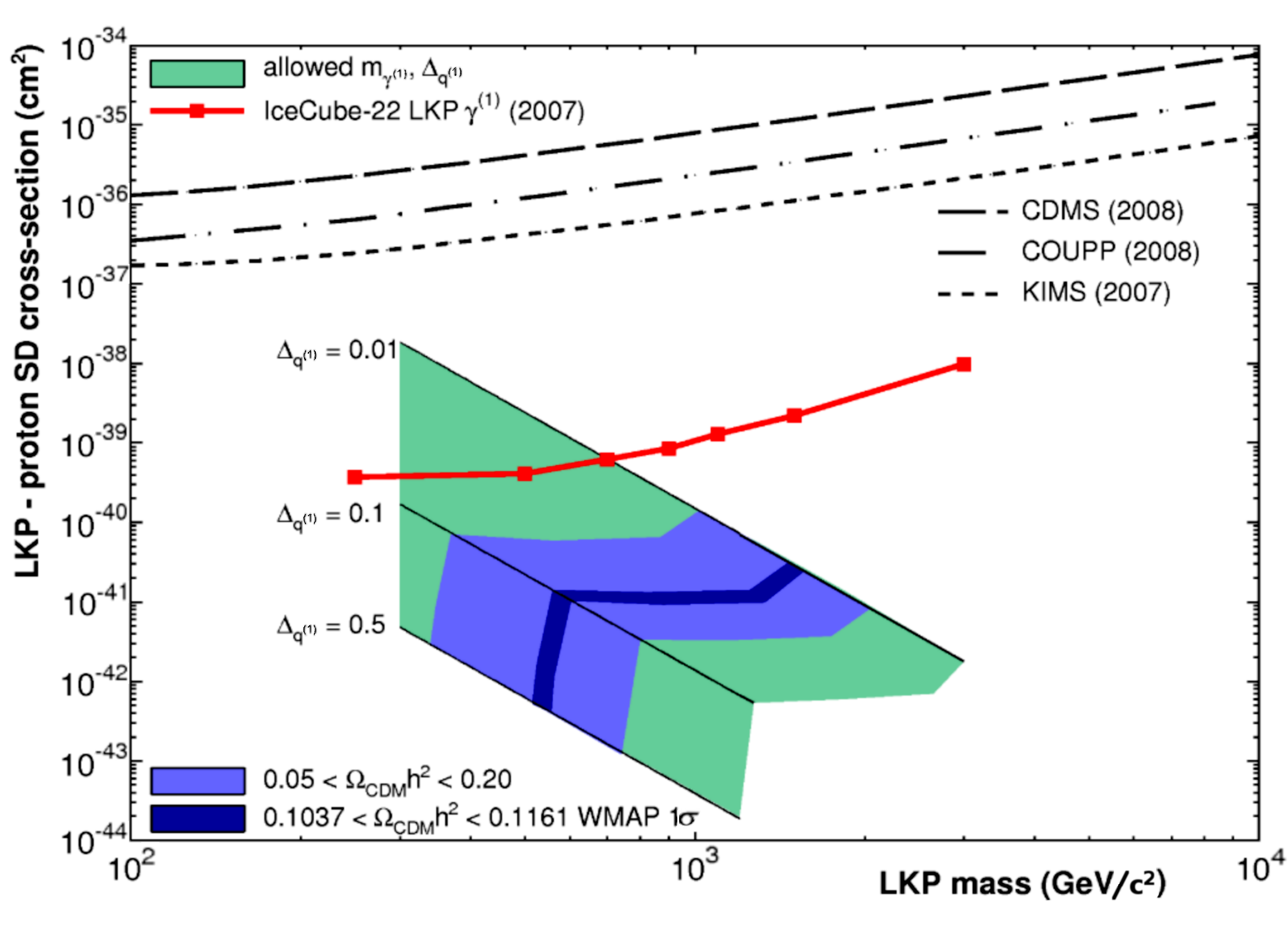}
\caption{
Limits on the spin-dependent cross-section of the lightest Kaluza-Klein
particle (LKP) and the region allowed in minimal models of extra dimensions. 
$\Delta_{q^{(1)}}$ is the relative mass splitting between the first
quark excitation and the LKP. 
The dark narrow (blue) region is obtained when consistency with 
the relic density inferred from WMAP measurements~\cite{WMAP}
is required within one sigma.
\label{fig:IC22KKlim}
}
\end{figure}

The negative result of the search in the IceCube 22-string data
was also interpreted~\cite{IC22KK} in the
framework of Kaluza-Klein theories with universal 
extra dimensions, 
with the WIMP annihilating in the Sun being ${\rm B}^{(1)}$,
the first excitation of the weak hypercharge gauge boson.
Figure \ref{fig:IC22KKlim} shows the cross-section limit obtained, and the
region allowed in minimial models. The limit from the seven-year AMANDA 
point source sample is almost identical.

\section{Future \label{sec:future}}

There are several initiatives to benefit from the presence of the IceCube
array and extend the range of physics topics which can be addressed.

The IceCube volume is too small to collect a sizeable sample of neutrino events
at EHE energies, because of the low flux. Acoustic and radio emissions 
from neutrino interactions are attenuated much less than light, and 
could be used to increase the effective volume at high energies in
a hybrid approach~\cite{hybrid}.

The South Pole Acoustic Test Setup~\cite{SPATS} has been used to study 
sound propagation and the acoustic environment in the ice using pingers
and receivers deployed in IceCube holes. The preliminary results indicate
an attenuation length of around $300\, {\rm m}$, which is more than an
order of magnitude below the predicted value~\cite{PriceAttenuation}.

Within IceCube there has also been a successful development programme, AURA, 
to test methods for radio detection in the ice~\cite{LandsmanAURA}.  
A proposal for an $\sim 80 {\rm km}^2$ array (Askaryan Radio Array, ARA) 
has been submitted. In addition, the possibility of a surface radio array to 
detect air showers was studied~\cite{AirRadio}.

At the other end of the energy scale, WIMP annihilations would probably
give neutrinos in the energy range from GeV to TeV. 
The original IceCube design is not optimal for these energies. Additional 
funding was secured, however, for six strings added
at the centre of IceCube. These are equipped with PMTs with 
enhanced quantum efficiency, about 30\% higher than that of standard
DOMs. The modules are placed closer to each other than on standard 
strings (7m as compared to 17m).
The instrumented region is between $2100\, {\rm m}$ and $2450\, {\rm m}$ (below 
the main dust layer), where the exceptionally clear ice, together with
the denser instrumentation, contributes to an overall ten-fold increase 
in photon collection efficiency. The lower portions of the central
thirteen strings, including the aforementioned six, constitute the
{\em Deep Core} subarray~\cite{DeepCore}, with a fiducial mass of about 13 MT. 
It is surrounded by at least three layers of
instrumentation on all sides. Simulations indicate
that an algorithm rejecting events with early hits outside the Deep Core
will be able to reduce the atmospheric muon background by a factor of
$10^6$, bringing it down to the level of atmospheric neutrinos~\cite{DCveto}.

\begin{figure}
\includegraphics[width=9cm]{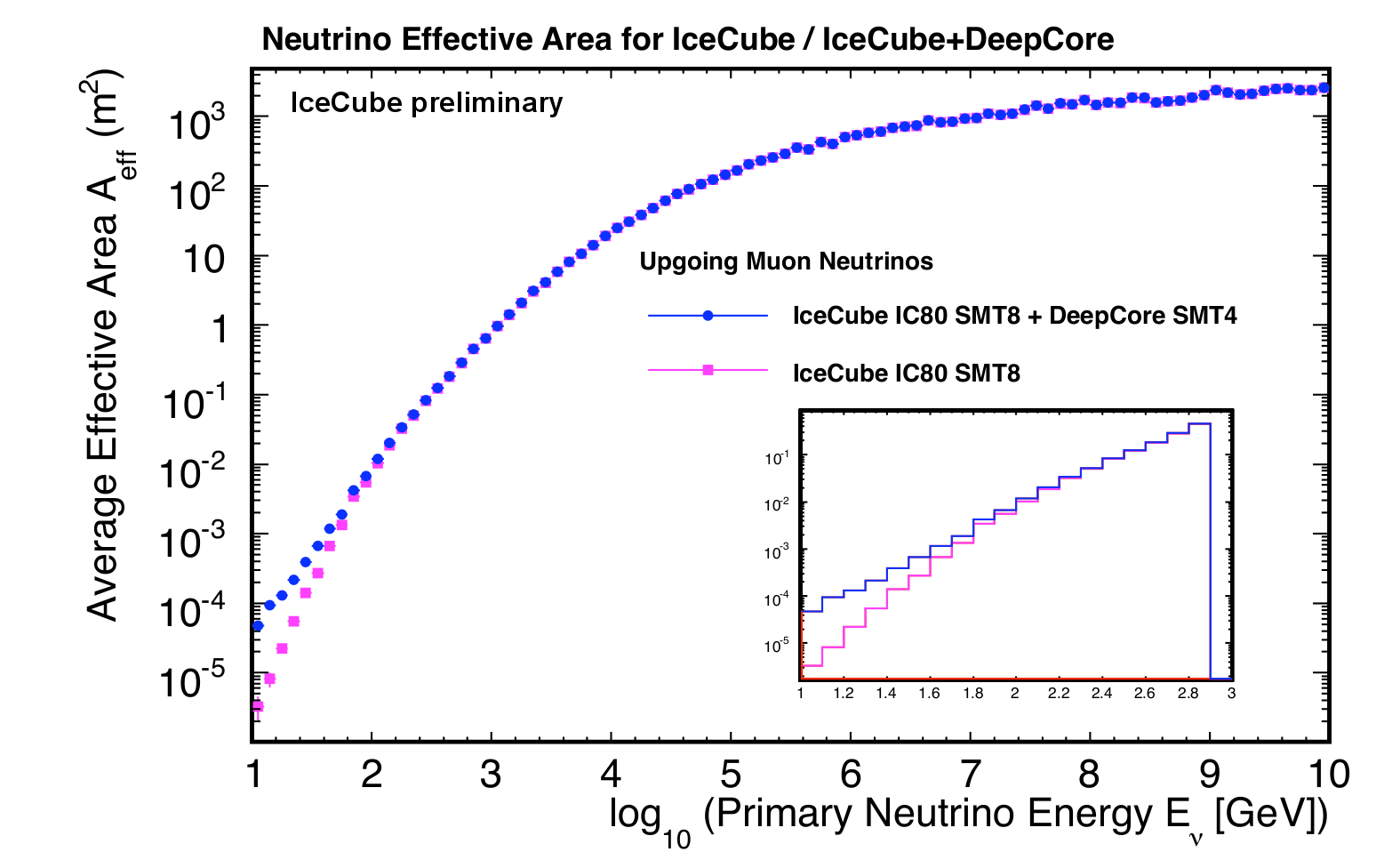}
\caption{
The effective area for muon neutrino detection at trigger level of 
the originally planned 80-string array, and with the addition 
of the Deep Core\label{fig:DCeffarea}
}
\end{figure}

Figure \ref{fig:DCeffarea} shows the IceCube effective area with and without 
the additional Deep Core strings. The improvement is significant at low energies.
The thick dotted line in Figure \ref{fig:NeutralinoSigma} shows the expected 
sensitivity in the solar WIMP search ten years after the completion of 
IceCube, using data taken with the Sun below the horizon. The sensitivity below 
$100\, {\rm GeV}/c^2$ is almost exclusively due to the Deep Core. 

The physics potential of the Deep Core goes well beyond the search for
Dark Matter~\cite{DCphysics}. The possibility to reject atmospheric muons 
entering through the surrounding detector array will make it possible to 
study neutrino emission from the southern hemisphere without requiring 
PeV energies. Furthermore, the atmospheric neutrino spectrum in the range 
between 10 GeV and 1 TeV will be accessible, which will make it possible to
conduct high-statistics studies of neutrino oscillations at $ \sim 10\, {\rm GeV}$ 
over baselines comparable to the size of the Earth. It will be
possible to study $\nu_\mu$ disappearance at higher energies than previously
done, and possibly $\nu_\tau$ appearance. It has also been suggested that
the Deep Core might have some sensitivity to the neutrino mass 
hierarchy~\cite{NuMassHierarchy}.

The additional six Deep Core strings were successfully installed during 
the 2009/2010 deployment season together with 14 standard IceCube strings,
bringing the total to 79 strings.

\section{Acknowledgements}
We acknowledge the support from the following agencies: 
U.S. National Science Foundation-Office of Polar Program, 
U.S. National Science Foundation-Physics Division, 
University of Wisconsin Alumni Research Foundation, 
U.S. Department of Energy, and National Energy Research Scientific Computing Center, the Louisiana Optical Network Initiative (LONI) grid computing resources; 
Swedish Research Council, 
Swedish Polar Research Secretariat, Swedish National Infrastructure for Computing,
and Knut and Alice Wallenberg Foundation, Sweden; 
German Ministry for Education and Research (BMBF), Deutsche Forschungsgemeinschaft (DFG), Research Department of Plasmas with Complex Interactions (Bochum), Germany; 
Fund for Scientific Research (FNRS-FWO), FWO Odysseus programme, 
Flanders Institute to encourage scientific and technological research in industry (IWT), 
Belgian Federal Science Policy Office (Belspo); 
Marsden Fund, New Zealand.




\bibliographystyle{elsarticle-num}



\end{document}